# Transferable E(3) equivariant parameterization for Hamiltonian of molecules and solids

Yang Zhong[1,2], Hongyu Yu[1,2], Mao Su[1,2], Xingao Gong[1,2], Hongjun Xiang[1,2*]

[1]Key Laboratory of Computational Physical Sciences (Ministry of Education), Institute of Computational Physical Sciences, State Key Laboratory of Surface Physics, and Department of Physics, Fudan University, Shanghai, 200433, China
[2]Shanghai Qi Zhi Institute, Shanghai, 200030, China
*E-mail: hxiang@fudan.edu.cn

## Abstract

Using the message-passing mechanism in machine learning (ML) instead of self-consistent iterations to directly build the mapping from structures to electronic Hamiltonian matrices will greatly improve the efficiency of density functional theory (DFT) calculations. In this work, we proposed a general analytic Hamiltonian representation in an E(3) equivariant framework, which can fit the ab initio Hamiltonian of molecules and solids by a complete data-driven method and are equivariant under rotation, space inversion, and time reversal operations. Our model reached state-of-the-art precision in the benchmark test and accurately predicted the electronic Hamiltonian matrices and related properties of various periodic and aperiodic systems, showing high transferability and generalization ability. This framework provides a general transferable model that can be used to accelerate the electronic structure calculations on different large systems with the same network weights trained on small structures.

## Introduction

Nowadays machine learning (ML) has a wide range of applications in molecular and materials science, including the direct prediction of various properties of materials[1-3], the construction of machine learning force fields (MLFFs) with quantum mechanical precision[4-7], the high-throughput generation of molecular and crystal structures[8-10], and the construction of more precise exchange-correlation functionals[11, 12]. However, the acquisition of the electronic structure of materials still relies almost exclusively on density functional theory (DFT) based calculations. Unfortunately, these methods are very time-consuming to get the Hamiltonian of the systems through self-consistent iterations and scale poorly with the system sizes. Semi-empirical tight-binding (TB) approximations[13], such as the Slater-Koster method[14], can reduce a lot of computation compared to the ab initio DFT methods. However, this approach often directly uses the existing or manually fine-tuned TB parameters and thus cannot accurately reproduce the electronic structure of general systems. Developing truly transferable, fully data-driven TB models applicable across materials, geometries, and boundary conditions can reconcile accuracy with speed but is rather challenging.

Hegde and Brown first used kernel ridge regression (KRR) to learn semi-empirical tight-binding Hamiltonian matrices[15]. They successfully fitted the Hamiltonian of the Cu system containing only rotation invariant $s$ orbitals and the diamond system consisting only of $s$ and $p$ orbitals. Similarly, Wang et al. designed a neural network model to obtain semi-empirical TB parameters by fitting the ab initio band structure[16]. An important feature of the Hamiltonian matrix is that its components transform equivariantly with the rotation of the coordinate system. However, none of these approaches deals with the rotational equivariance of the Hamiltonian matrix. Moreover, the two methods can only fit the empirical model Hamiltonian rather than the true ab initio tight-binding Hamiltonian matrices generated by the self-consistent iteration of ab initio tight-binding methods such as OpenMX[17, 18] and Siesta[19, 20].

Zhang et al.[21] and Nigam et al.[22] proposed a method to predict the ab-initio TB Hamiltonian of small molecules and simple solid systems by constructing an equivariant kernel in Gaussian Process Regression (GPR) to parameterize the Hamiltonian. Since GPR uses a fixed kernel and representation[23-26], its training accuracy and multi-element generalization ability are usually lower than those of deep neural networks such as the message passing neural networks (MPNNs)[27-32] when the number of training samples is sufficient. Therefore, developing graph neural networks (GNNs) capable of predicting the Hamiltonian of general periodic and aperiodic systems would be the best option.

However, traditional GNNs can only predict rotation-invariant scalars such as energy, band gap, etc. GNNs must encode the directional information of the system in an appropriate way to predict equivariant directional properties. To make the predicted Hamiltonian matrices satisfy the rotational equivariance, Schütt et al. designed the SchNorb neural network architecture by embedding the direction information of the bonds into the message-passing function[33]. This network constructs the ab initio Hamiltonian of molecules from the directional edge features of atom pairs. However, SchNorb needs to learn the rotational equivariance of the Hamiltonian matrix through data augmentation, which greatly increases the amounts of training data and redundant parameters of the network. Unke et al. proposed the PhisNet model[34], which realized the SE(3) equivariant parameterization of the Hamiltonian matrices with GNN based on SO(3) representations and achieved state-of-the-art accuracy on the Hamiltonian of small molecules such as water and ethanol. However, it should be noted that the PhisNet model is not the most universal representation of the Hamiltonian as it ignores the parity of the Hamiltonian, which may lead to serious problems when predicting periodical systems with infinite sizes.

Recently, Li et al. proposed a GNN model called DeepH to predict the ab initio Hamiltonian by constructing a local coordinate system in a crystal[35]. DeepH successfully predicted the tight-binding Hamiltonian of some simple periodic systems such as graphene, carbon nanotubes, etc. Their original intention of introducing the local coordinate system is to solve the rotation equivariance problem of the Hamiltonian, but DeepH still embeds the local directional information of interacting atom pairs in the invariant message passing function, which will undoubtedly increase the number of redundant parameters of the network but may require less data augmentation than a fully invariant model without local coordinate systems. In addition, the hopping distance between two interacting atoms far exceeds the lengths of the general chemical bonds. Taking the smallest hydrogen atom as an example, the cutoff radius of the numerical atomic orbital of the hydrogen atom used by OpenMX is 6 Bohr, so the furthest hopping between any two atomic bases used by OpenMX in periodic systems can exceed at least 12 Bohr (~6.4 Å), a distance that even exceeds the lattice parameters of some crystals. Therefore, it is difficult to describe such long hopping in a well-defined local coordinate system.

Because of the spherical harmonic part of the atomic basis functions, the TB Hamiltonian matrix must satisfy two fundamental constraints: rotational equivariance and parity symmetry. When the spin-orbit coupling (SOC) effects or the ionic magnetic moments are taken into account, rotational equivariance in the spin degrees of freedom and additional time-reversal equivariance need to be fulfilled. It is hard for the models to learn the physically correct dependence on the direction of input structures from the data.

In this work, we constructed a general parametrized Hamiltonian by decomposing each block of the Hamiltonian into a vector coupling of equivariant irreducible spherical tensors[36] (ISTs) with correct parity symmetry. This parametrized Hamiltonian strictly satisfies the rotational equivariance and parity symmetry and can be extended to a parameterized Hamiltonian satisfying SU(2) and time-reversal equivariance to fit the Hamiltonian with SOC effects or ionic magnetic moments. Based on this universal parametrized Hamiltonian, we designed the E(3) equivariant HamGNN model for predicting the ab initio TB Hamiltonian of molecules and solids. HamGNN has reached state-of-the-art accuracy on the benchmark test and shows high efficiency and transferability in the prediction of various periodic and aperiodic systems. The trained HamGNN model can predict the Hamiltonian matrices, energy bands, and wavefunctions of the structures not present in the training set. The high transferability and precision of our model enable this ML electronic structure method to replicate the success of MLFFs and be widely used in practical electronic structure calculations.

## Results

### *E(3) equivariant parametrized Hamiltonian*

The core of the electronic structure problem in DFT is to solve the Kohn-Sham equation for electrons in reciprocal space. If the Kohn-Sham Hamiltonian is represented by numerical atomic orbitals centered on each atom (such as those defined in the OpenMX and Siesta packages), then the Kohn-Sham equation can be expressed as a generalized eigenvalue problem as follows:

$$\mathbf{H}^{(\vec{k})} \psi_{n\vec{k}} = \varepsilon_{n\vec{k}} \mathbf{S}^{(\vec{k})} \psi_{n\vec{k}}, \tag{1}$$

where $\mathbf{H}^{(\vec{k})}_{n_i l_i m_i, n_j l_j m_j} = \sum_{n_c} e^{i\vec{k}\cdot\vec{R}_{n_c}} \mathbf{H}^{(\vec{R}_{n_c})}_{n_i l_i m_i, n_j l_j m_j}$ and $\mathbf{S}^{(\vec{k})}_{n_i l_i m_i, n_j l_j m_j} = \sum_{n_c} e^{i\vec{k}\cdot\vec{R}_{n_c}} \mathbf{S}^{(\vec{R}_{n_c})}_{n_i l_i m_i, n_j l_j m_j}$ are the Kohn-

Sham Hamiltonian and overlap matrices at the point $\vec{k}$ in the reciprocal space. $\mathbf{H}^{(\vec{k})}_{n_i l_i m_i, n_j l_j m_j}$ and $\mathbf{S}^{(\vec{k})}_{n_i l_i m_i, n_j l_j m_j}$ are obtained by Fourier transform of real-space TB Hamiltonian matrix $\mathbf{H}^{(\vec{R}_{n_c})}_{n_i l_i m_i, n_j l_j m_j} = \left\langle \phi_{n_i l_i m_i}(\vec{r} - \vec{\tau}_i) \middle| \hat{H} \middle| \phi_{n_j l_j m_j}(\vec{r} - \vec{\tau}_j - \vec{R}_{n_c}) \right\rangle$ and overlap matrix $\mathbf{S}^{(\vec{R}_{n_c})}_{n_i l_i m_i, n_j l_j m_j} = \left\langle \phi_{n_i l_i m_i}(\vec{r} - \vec{\tau}_i) \middle| \phi_{n_j l_j m_j}(\vec{r} - \vec{\tau}_j - \vec{R}_{n_c}) \right\rangle$ in the basis of atomic orbitals $\phi_{n_i l_i m_i}$ at the site $\vec{\tau}_i$ and $\phi_{n_j l_j m_j}$ at the site $\vec{\tau}_j + \vec{R}_{n_c}$, where $\vec{R}_{n_c}$ is the shift vector of periodic image cell. Therefore, once we have obtained the Hamiltonian matrix and overlap matrix in real space, we can further solve the electronic structure in the whole reciprocal space.

Due to the spherical symmetry of the atomic potential, the atomic orbital bases as its eigenfunctions not only satisfy the rotational equivariance under the operation $Q \in SO(3)$ but also has a certain parity symmetry under the inversion operation $g \in \{E, I\}$. Under a rotatory inversion operation $gQ \in O(3)$, the TB Hamiltonian matrix element $\mathbf{H}_{n_i l_i m_i, n_j l_j m_j}$ in real space becomes (we omit the notation $\vec{R}_{n_c}$ for convenience in the following discussion):

$$\mathbf{H}'_{n_i l_i m_i, n_j l_j m_j} = \left\langle gQ\phi_{n_i l_i m_i} \middle| \hat{H} \middle| gQ\phi_{n_j l_j m_j} \right\rangle. \tag{2}$$

The irreducible representation of $gQ$ is $\sigma_p(g) \otimes D(Q)$, where $D(Q)$ is the Wigner $D$ matrix and $\sigma_p(g)$ is the scalar irreducible representation of the inversion operation, which is defined as follows

$$\sigma_p(g) = \begin{cases} 1, & g = E \\ p, & g = I \end{cases}. \tag{3}$$

Substitute the irreducible representation of $gQ$ into Eq. (2), we can get

$$\mathbf{H}'_{n_i l_i m_i, n_j l_j m_j} = \sigma_{p_i p_j}(g) \sum_{\mu_i \mu_j} D^{(l_i)}_{m_i \mu_i}(Q) D^{(l_j)}_{m_j \mu_j}(Q) \mathbf{H}_{n_i l_i \mu_i, n_j l_j \mu_j}, \tag{4}$$

where $\sigma_{p_i p_j}(g) = \sigma_{p_i}(g)\sigma_{p_j}(g)$. We further write the right-hand side of Eq. (4) in the form of matrix-vector multiplication:

$$\mathbf{H}'_{n_i l_i m_i, n_j l_j m_j} = \sigma_{p_i p_j}(g) \sum_{\mu_i \mu_j} \left[ D^{l_i}(Q) \otimes D^{l_j}(Q) \right]_{m_i m_j, \mu_i \mu_j} \mathbf{H}_{n_i l_i \mu_i, n_j l_j \mu_j}. \tag{5}$$

It can be seen from the above equation that each sub-block $\mathbf{H}_{n_i l_i \mu_i, n_j l_j \mu_j}$ ($|\mu_i| \leq l_i, |\mu_j| \leq l_j$) of the TB Hamiltonian based on atomic orbitals can be regarded as a spherical tensor[36,37] $\boldsymbol{T}^{n_i l_i, n_j l_j}_{\boldsymbol{\mu}, p_i p_j} \equiv \mathbf{H}_{n_i l_i \mu_i, n_j l_j \mu_j}$ with the parity $p_i p_j$, which is rotationally equivariant according to the generalized Wigner D matrix $\boldsymbol{D}^{\boldsymbol{l}}_{\boldsymbol{\mu m}}(Q) = D^{l_i}_{\mu_i m_i}(Q) D^{l_j}_{\mu_j m_j}(Q)$, where $\boldsymbol{l} = (l_i, l_j)$, $\boldsymbol{\mu} = (\mu_i, \mu_j)$, $\boldsymbol{m} = (m_i, m_j)$.

According to the angular momentum theory[37,38], $D^{l_i}(Q) \otimes D^{l_j}(Q)$ is a reducible representation, which can be further decomposed into the direct sum of several irreducible Wigner D matrices:

$$D^{l_i} \otimes D^{l_j} = D^{|l_i - l_j|} \oplus D^{|l_i - l_j|+1} \oplus \cdots \oplus D^{l_i + l_j}. \tag{6}$$

Combining the parity of the Hamiltonian matrix block $(n_i l_i, n_j l_j)$, we can get

$$\sigma_{p_i p_j}(g) D^{l_i}(Q) \otimes D^{l_j}(Q) = \sum_{L=|l_i-l_j|}^{l_i+l_j} \oplus \sigma_{p_i p_j}(g) D^L(Q). \tag{7}$$

According to Eq. (7), $\boldsymbol{T}^{n_i l_i, n_j l_j}_{\boldsymbol{\mu}, p_i p_j}$ is reducible and the coupled irreducible spherical tensor $T^{n_i l_i, n_j l_j}_{L, p_i p_j, m}$ in each order $L = |l_i - l_j|, \cdots, l_i + l_j$ can be obtained by the vector coupling of $\boldsymbol{T}^{n_i l_i, n_j l_j}_{\boldsymbol{\mu}, p_i p_j}$:

$$T^{n_i l_i, n_j l_j}_{L, p_i p_j, m} = \sum_{\mu_i = -l_i}^{l_i} \sum_{\mu_j = -l_j}^{l_j} C^{L l_i l_j}_{m \mu_i \mu_j} \boldsymbol{T}^{n_i l_i, n_j l_j}_{\boldsymbol{\mu}, p_i p_j}, \tag{8}$$

where $C^{L l_i l_j}_{m \mu_i \mu_j}$ is the vector coupling coefficient, namely the Clebsch-Gordan coefficient. Each IST $T_{L,p_ip_j,m}^{n_il_i,n_jl_j}$ has the parity symmetry of $p_ip_j$ and satisfies the rotational equivariance of order $L$. By inverse linear transformation of Eq. (8), $\boldsymbol{T}_{\boldsymbol{\mu},p_ip_j}^{n_il_i,n_jl_j}$ can be constructed from ISTs $T_{L,p_ip_j,m}^{n_il_i,n_jl_j}$:

$$\boldsymbol{T}_{\boldsymbol{\mu},p_ip_j}^{n_il_i,n_jl_j} = \sum_{L=|l_i-l_j|}^{l_i+l_j} \sum_{m=-L}^{L} C_{\mu_i\mu_j m}^{l_il_jL} T_{L,p_ip_j,m}^{n_il_i,n_jl_j}. \tag{9}$$

Therefore, as long as we find all ISTs corresponding to each block of the Hamiltonian matrix, we can construct the entire Hamiltonian matrix in a block-wise manner through Eq. (9). We construct two O(3) equivariant vectors $\boldsymbol{\Omega}_i^{on}$ and $\boldsymbol{\Omega}_{ij}^{off}$ by direct summation of all the ISTs required by the on-site ($i = j$) Hamiltonian and the off-site ($i \neq j$) Hamiltonian respectively:

$$\boldsymbol{\Omega}_i^{on} = \sum_{n_il_i} \sum_{n_i'l_i'} \sum_{L=|l_i-l_i'|}^{l_i+l_i'} \oplus \left[ T_{L,p_ip_i',m}^{n_il_i,n_i'l_i'} \right]_{-L\leq m\leq L}, \tag{10}$$

$$\boldsymbol{\Omega}_{ij}^{off} = \sum_{n_il_i} \sum_{n_jl_j} \sum_{L=|l_i-l_j|}^{l_i+l_j} \oplus \left[ T_{L,p_ip_j,m}^{n_il_i,n_jl_j} \right]_{-L\leq m\leq L}. \tag{11}$$

The prediction of the Hamiltonian is transformed into the prediction of $\boldsymbol{\Omega}_i^{on}$ and $\boldsymbol{\Omega}_{ij}^{off}$, which can be obtained by mapping from the equivariant features of the nodes and the pair interactions, respectively. The final parameterized Hamiltonian can be expressed as:

$$\tilde{H}_{n_il_im_i,n_jl_jm_j} = \begin{cases} \tilde{H}_{n_il_im_i,n_i'l_i'm_i'}^{on} = \sum_{L=|l_i-l_i'|}^{l_i+l_i'} \sum_{m=-L}^{L} C_{m_i,m_i',m}^{l_i,l_i',L} T_{L,p_ip_i',m}^{n_il_i,n_i'l_i'} & i=j \\ \tilde{H}_{n_il_im_i,n_jl_jm_j}^{off} = \sum_{l_3=|l_i-l_j|}^{l_i+l_j} \sum_{m=-L}^{L} C_{m_i,m_j,m}^{l_i,l_j,L} T_{L,p_ip_j,m}^{n_il_i,n_jl_j} & i\neq j \end{cases}. \tag{12}$$

The above formula is O(3) equivariant. Since GNN naturally has translational symmetry, the parameterized Hamiltonian represented by $\boldsymbol{\Omega}_i^{on}$ and $\boldsymbol{\Omega}_{ij}^{off}$ obtained from GNN has E(3) equivariance.

When the spin-orbit coupling (SOC) effects are considered, the real-space Hamiltonian matrices are complex-valued and can be divided into four sub-blocks $\hat{\mathbf{H}}_{s_i s_j}$ ($s_i, s_j = \uparrow \text{ or } \downarrow$) by the spin degree. In this case, the complete Hamiltonian matrices satisfy the following SU(2) rotational equivariance:

$$\left\langle gQ(\phi_{n_i l_i m_i} s_i) \middle| \hat{\mathbf{H}} \middle| gQ(\phi_{n_j l_j m_j} s_j) \right\rangle = \sigma_{p_i p_j}(g) \sum_{\mu_i=-l_i}^{l_i} \sum_{\mu_j=-l_j}^{l_j} \sum_{s'_i=-1/2}^{1/2} \sum_{s'_j=-1/2}^{1/2} \left( D_{m_i \mu_i}^{(l_i)}(Q) D_{m_j \mu_j}^{(l_j)}(Q) \right.$$

$$\left. D_{s_i s'_i}^{(1/2)*}(Q) D_{s_j s'_j}^{(1/2)}(Q) \left\langle \phi_{n_i l_i \mu_i} s'_i \middle| \hat{\mathbf{H}} \middle| \phi_{n_j l_j \mu_j} s'_j \right\rangle \right). \tag{13}$$

Although each subblock $\hat{\mathbf{H}}_{s_i s_j}$ satisfies the O(3) rotational equivariance, they are coupled to each other under the rotational operations. Therefore, the four subblocks predicted independently with the O(3) equivariant parameterized Hamiltonian cannot be used to construct the complete SU(2) equivariant Hamiltonian with the SOC effect. In addition, the real and imaginary parts of the SU(2) equivariant Hamiltonian matrices are also coupled during rotation, so the complete Hamiltonian cannot be constructed by using the independently predicted real and imaginary parts. These methods not only rely on a large number of fitting parameters but also can not make the constructed Hamiltonian matrices strictly meet the SU(2) equivariance. To ensure that the SOC effect learned by the network complies with the physical rules and SU(2) equivariance, we explicitly express the complete Hamiltonian as the sum of the spin-less part and the SOC part:

$$\hat{\mathbf{H}} = \hat{H} \otimes I_2 + \hat{H}_{SOC}, \tag{14}$$

where $\hat{H}_{SOC} = \frac{1}{2} \xi(r) \hat{\vec{L}} \cdot \hat{\vec{\sigma}} = \frac{1}{2} \xi(r) \left( \hat{L}_x \hat{\sigma}_x + \hat{L}_y \hat{\sigma}_y + \hat{L}_z \hat{\sigma}_z \right)$, which satisfies SU(2) rotational equivariance and time-reversal equivariance. $\xi(r)$ is an invariant coefficient describing the strength of the SOC effects[39]. According to the above equation, we can get the following parameterized Hamiltonian matrices with SOC effect in the atomic orbitals[39, 40]:

$$\tilde{H}^{SOC}_{n_i l_i m_i s_i, n_j l_j m_j s_j} = \begin{cases} \tilde{H}_{n_i l_i m_i, n_j l_j m_j} + \frac{1}{2}\xi_{n_i l_i, n_j l_j}\left\langle \phi_{n_i l_i m_i} \middle| \hat{L}_z \middle| \phi_{n_j l_j m_j} \right\rangle & s_i = \uparrow, s_j = \uparrow \\ \frac{1}{2}\xi_{n_i l_i, n_j l_j}\left(\left\langle \phi_{n_i l_i m_i} \middle| \hat{L}_x \middle| \phi_{n_j l_j m_j} \right\rangle - i\left\langle \phi_{n_i l_i m_i} \middle| \hat{L}_y \middle| \phi_{n_j l_j m_j} \right\rangle\right) & s_i = \uparrow, s_j = \downarrow \\ \frac{1}{2}\xi_{n_i l_i, n_j l_j}\left(\left\langle \phi_{n_i l_i m_i} \middle| \hat{L}_x \middle| \phi_{n_j l_j m_j} \right\rangle + i\left\langle \phi_{n_i l_i m_i} \middle| \hat{L}_y \middle| \phi_{n_j l_j m_j} \right\rangle\right) & s_i = \downarrow, s_j = \uparrow \\ \tilde{H}_{n_i l_i m_i, n_j l_j m_j} - \frac{1}{2}\xi_{n_i l_i, n_j l_j}\left\langle \phi_{n_i l_i m_i} \middle| \hat{L}_z \middle| \phi_{n_j l_j m_j} \right\rangle & s_i = \downarrow, s_j = \downarrow \end{cases}. \quad (15)$$

Since the matrix representation of the angular momentum operator $\left\langle \phi_{n_i l_i m_i} \middle| \hat{L}_\alpha \middle| \phi_{n_j l_j m_j} \right\rangle = \left\langle \phi_{n_i l_i m_i}(\vec{r}_i) \middle| \hat{L}_\alpha \middle| \phi_{n_j l_j m_j}(\vec{r}_j - \vec{\tau}_{ji}) \right\rangle$ under the atomic orbital basis can be directly calculated analytically, the only learnable parameters in Eq. (15) are $\tilde{H}_{n_i l_i m_i, n_j l_j m_j}$ and $\xi_{n_i l_i, n_j l_j}$. $\tilde{H}_{n_i l_i m_i, n_j l_j m_j}$ can be directly expressed by Eq. (12), and $\xi_{n_i l_i, n_j l_j}$ is an invariant scalar coefficient that can be mapped from the features of atom pairs $ij$.

We can further derive the parameterized Hamiltonian satisfying the time-reversal equivariance for the magnetic systems. The classical Heisenberg model can be written in a general matrix form containing all possible second-order interactions[41]:

$$E_{spin} = \sum_{\langle i',j' \rangle} \sum_{\alpha\beta} J^{\alpha\beta}_{i'j'}\left(M^{\alpha}_{i'}[\rho] \cdot M^{\beta}_{j'}[\rho]\right) + \sum_{k'} \sum_{\alpha\beta} A^{\alpha\beta}_{k'}\left(M^{\alpha}_{k'}[\rho] \cdot M^{\beta}_{k'}[\rho]\right), \quad (16)$$

where the 3×3 tensors $J^{\alpha\beta}_{i'j'}$ and $A^{\alpha\beta}_{k'}$ are called the $J$ matrix and single-ion anisotropy (SIA) matrix, respectively. $M^{\alpha}_{i'}[\rho]$ is the component of the ionic magnetic moment $\vec{M}_{i'}[\rho] = Tr\left[\rho \hat{W}_{i'} \hat{\vec{\sigma}}\right]$, which is a functional of the electron density $\rho$. The magnetization of the system is partitioned into the local magnetic moments on each atom by a pre-defined weight operator $\hat{W}_{i'}$, which is commonly defined as $\hat{W}_{i'} = \int d\vec{r} f\left(r_{cut} - \|\vec{r} - \vec{\tau}_{i'}\|\right)|\vec{r}\rangle\langle\vec{r}|$, where $f$ is a radial cutoff function centered on atom $i'$. The matrix element of the weight operator $\hat{W}_{i'}$ is

given by[42, 43]

$$\left(\hat{W}_{i'}\right)_{n_i l_i m_i, n_j l_j m_j} = \begin{cases} w_{n_i l_i m_i, n_j l_j m_j}, & i, j = i' \\ \frac{1}{2} w_{n_i l_i m_i, n_j l_j m_j}, & i \text{ or } j = i' \\ 0, & i, j \neq i' \end{cases} \quad (17)$$

The matrix $w_{n_i l_i m_i, n_j l_j m_j}$ varies with the choice of the cutoff function $f$ and can be parameterized by Eq. (12). The magnetic part of the parameterized Hamiltonian for the magnetic systems equals the variational derivative $\frac{\delta E_{spin}}{\delta \rho}$:

$$\tilde{H}^{mag}_{n_i l_i m_i s_i, n_j l_j m_j s_j} = \sum_{\langle i', j' \rangle} \sum_{\alpha \beta} J^{\alpha \beta}_{i'j'} \left( M^{\alpha}_{i'} \left(W_{j'}\right)_{n_i l_i m_i, n_j l_j m_j} \sigma_{\beta, s_i s_j} + M^{\beta}_{j'} \left(W_{i'}\right)_{n_i l_i m_i, n_j l_j m_j} \sigma_{\alpha, s_i s_j} \right) +$$

$$\sum_{k'} \sum_{\alpha \beta} A^{\alpha \beta}_{k'} \left(W_{k'}\right)_{n_i l_i m_i, n_j l_j m_j} \left( M^{\alpha}_{k'} \sigma_{\beta, s_i s_j} + M^{\beta}_{k'} \sigma_{\alpha, s_i s_j} \right). \quad (18)$$

The $J^{\alpha\beta}_{i'j'}$ and $A^{\alpha\beta}_{k'}$ tensors are learnable and can be mapped from the features of the edges and nodes respectively. Since the spin magnetic moment is odd under time-reversal operation and even under spatial inversion operation, the tensors $J^{\alpha\beta}_{i'j'}$ and $A^{\alpha\beta}_{k'}$ should be even under time-reversal and spatial inversion operations so that the parameterized Hamiltonian matrix constructed by Eq. (18) satisfies the time-reversal equivariance and the parity symmetry. In addition, Eq. (18) still satisfies all the equivariance when only the ionic magnetic moments are rotated.

## *Network Implementation*

As can be seen from the above discussion, each Hamiltonian matrix block satisfies the rotational equivariance and has a definite parity under the inversion operation, so we designed E(3) equivariant HamGNN deep neural network based on MPNN to fit ab initio TB Hamiltonian. This framework directly captures the electronic structure without expensive self-consistent iterations by constructing local equivariant representations of each atomic orbit. The network architecture of HamGNN is shown in Fig. 1(a). HamGNN can achieve a direct mapping from atomic species $\{Z_i\}$ and positions $\{\vec{r}_i\}$ to ab initio TB Hamiltonian matrix.

HamGNN first encodes elements, inter-atomic distances, and relative directions as initial graph embeddings. The distance between atom $i$ and its neighboring atom $j$ within the cutoff radius $r_c$ is expanded using the Bessel basis function:

$$B(\|\vec{r}_{ij}\|) = \sqrt{\frac{2}{r_c}} \frac{\sin(n\pi \|\vec{r}_{ij}\|/r_c)}{\|\vec{r}_{ij}\|} f_c(\|\vec{r}_{ij}\|), \tag{19}$$

where $f_c$ is the cosine cutoff function, which guarantees physical continuity for the neighbor atoms close to the cutoff sphere. The directional information between atom $i$ and atom $j$ is embedded in a set of real spherical harmonics $\{Y_{m_f}^{l_f}(\hat{\vec{r}}_{ij})\}$, which is used to construct the rotation-equivariant filter[44] in the equivariant message passing.

The atomic feature $V = V_{l_0 p_0} \oplus V_{l_1 p_1} \oplus \cdots \oplus V_{l_{max} p_{max}}$ in HamGNN is represented by a direct sum of different O(3) representations up to $l_{max}$ order, and each order feature can characterize the atomic orbit with different angular quantum numbers. Such features are transformed by the direct sum of the Wigner $D$ matrix, i.e. by the block diagonal matrix $\boldsymbol{D} = D_{l_0} \oplus D_{l_1} \oplus \cdots \oplus D_{l_{max}}$. $V_{l_i, p_i, c_i, m_i}^{i,t}$ denotes the equivariant features of atom $i$ in the orbital convolution layer $t$, where $l_i \leq l_{max}$ is the order of the O(3) irreducible representation, $p_i \in \{1, -1\}$ denotes the parity of the equivariant components of the order $l_i$, $-l_i \leq m_i \leq l_i$ is the index of each projection of the equivariant representation, $c_i$ is the channel index. We use T orbital convolution layers to construct the equivariant features of the atomic orbits in the local environment. Each orbital convolution layer performs a tensor product of the equivariant features of the neighbor atomic orbits and the spherical harmonic embeddings of the edge directions via Eq. (20), and the invariant scalar features of the interatomic distances are used to scale the equivariant output of each angular momentum. The message $m_{l_i,p_i,c_i,m_i}^{ij,t}$ output by Eq. (20) is composed of contributions with parity $p_i = p_j p_f$ to satisfy parity symmetry. Eq. (21) is the update function of orbital equivariant features. The updated orbital equivariant features are passed to a nonlinear gate activation function[45], which scales the input features equivariantly with the invariant field ($l \neq 0$) of the input features as the gate.

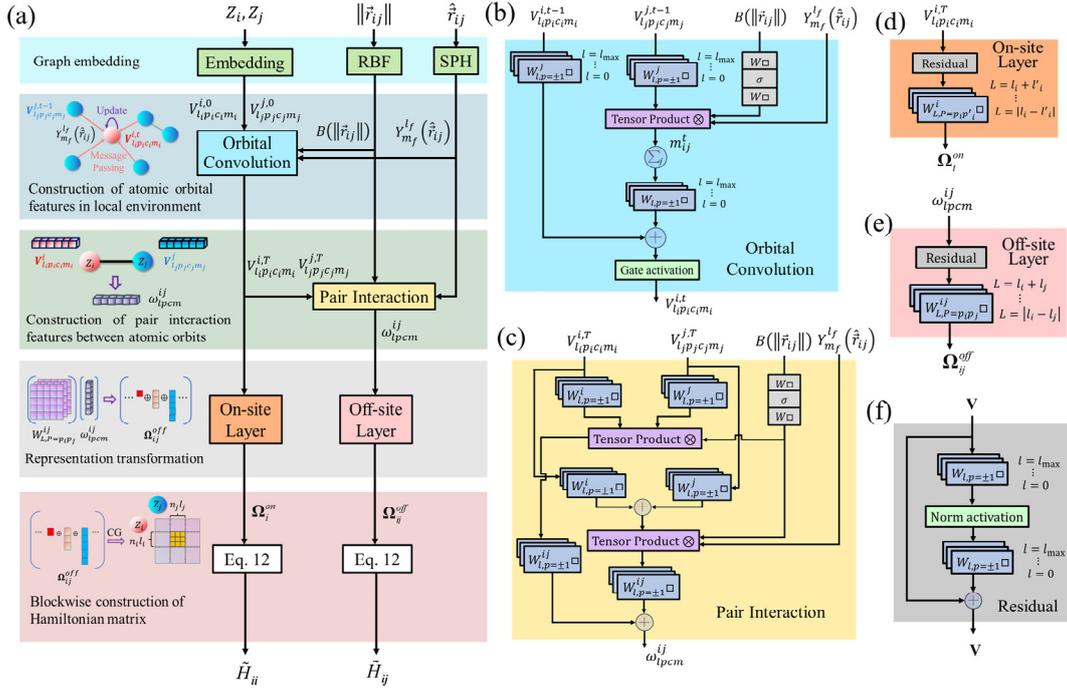

**Fig. 1. HamGNN architecture and the illustration of its subnetworks.** (a) The overall architecture of HamGNN. This neural network architecture predicts the Hamiltonian matrix through 5 steps. The prediction starts from the initial graph embedding of the species, interatomic distances, and interatomic directions of molecules and crystals. The atomic orbital features with angular momentum *l* in the local environment are included in the *l*-order components of the E(3) equivariant atom features and are refined through T orbital convolution blocks. In the third step, pair interaction features $\omega_{l,p,c,m}^{ij}$ of atomic orbitals are constructed by pair interaction blocks. In the fourth step, the IST representations of on-site and off-site Hamiltonian matrices are constructed by passing the features of atomic orbitals and pair interactions through the on-site layer and off-site layer, respectively. The final step is to construct the on-site and off-site Hamiltonian matrices block-by-block via the parameterized Hamiltonian given by Eq. (12). (b) Orbital convolution block. The equivariant atomic features that include the features of atomic orbitals of each angular momentum *l* are refined by the equivariant message passing and update functions. (c) Pair interaction block. This block is used to construct the pair interaction features between the orbitals of two adjacent atoms by equivariant tensor product. (d) On-site layer. The equivariant features of atoms are transformed into ISTs of on-site blocks $\Omega_i^{on}$ by the on-site layer. (e) Off-site layer. The pair interaction features between atomic orbitals are transformed into ISTs of the off-site block $\Omega_{ij}^{off}$ by the off-site layer. (f) Residual block. The residual block is used in the on-site layer and off-site layer to perform a nonlinear equivariant transformation of input features.

$$m_{l_i,p_i,c_i,m_i}^{ij,t} = \sum_{m_f=-l_f}^{l_f} \sum_{m_j=-l_j}^{l_j} C_{m_i,m_f,m_i}^{l_i,l_f,l_i} MLP\left[B(r_{ij})\right]_{l_f,p_f,l_j,p_j}^{l_i,p_i,c_i} Y_{m_f}^{l_f}(\hat{\vec{r}}_{ij}) V_{l_j,p_j,c_i,m_j}^{j,t-1} \quad (20)$$

$$V_{l_i,p_i,c_i,m_i}^{i,t} = V_{l_i,p_i,c_i,m_i}^{i,t-1} + \sum_{j \in N(i)} m_{l_i,p_i,c_i,m_i}^{ij,t} \quad (21)$$

The pair interaction layer adjusts the pair interaction features (used to construct off-site Hamiltonian) based on the features of the atomic orbits of two interacting atoms as well as the direction and strength of their interactions by the following equation:

$$\omega_{l,p,c,m}^{ij} = \sum_{m_i=-l_i}^{l_i} \sum_{m_j=-l_j}^{l_j} C_{m_i,m_j,m}^{l_i,l_j,l} MLP\left[B(r_{ij})\right]_{l_i,p_i,l_j,p_j}^{l,p,c} \left(\sum_{c'} W_{l_i,p_i,c,c'}^{i} V_{l_i,p_i,c',m_i}^{i,T}\right) \left(\sum_{c'} W_{l_j,p_j,c,c'}^{j} V_{l_j,p_j,c',m_j}^{j,T}\right)$$

$$+ \sum_{m_f=-l_f}^{l_f} \sum_{m'=-l'}^{l'} C_{m',m_f,m}^{l',l_f,l} MLP\left[B(r_{ij})\right]_{l_f,p_f,l',p'}^{l,p,c} Y_{m_f}^{l_f}(\hat{\vec{r}}_{ij}) V_{l',p',c,m'}^{ij}, \quad (22)$$

where $V_{l',p',c,m'}^{ij} = \sum_{c'} W_{l',p',c,c'}^{i} V_{l',p',c',m'}^{i,T} + \sum_{c'} W_{l',p',c,c'}^{j} V_{l',p',c',m'}^{j,T}$ is the mixed feature vector of $V_{l_i,p_i,c_i,m_i}^{i,T}$ and $V_{l_j,p_j,c_j,m_j}^{j,T}$. The on-site layer and off-site layer are used to convert the node features $V_{l_i,p_i,c_i,m_i}^{i,T}$ and pair interaction features $\omega_{l,p,c,m}^{ij}$ into the direct sums $\mathbf{\Omega}_i^{on}$ and $\mathbf{\Omega}_{ij}^{off}$ of the ISTs required to construct on-site and off-site Hamiltonian blocks, respectively. We add shortcut connections in the on-site layer and off-site layer and also use a norm activation function that scales the modulus of the irreducible representations of each order nonlinearly to increase the nonlinear fitting ability of the network. In the last step, the network uses the ISTs in $\mathbf{\Omega}_i^{on}$ and $\mathbf{\Omega}_{ij}^{off}$ to construct the on-site and off-site Hamiltonian blocks through Eq. (12).

The predicted Hermitian Hamiltonian is obtained by the following symmetrization:

$$H_{n_i l_i m_i, n_j l_j m_j} = \begin{cases} H^{on}_{n_i l_i m_i, n'_i l'_i m'_i} = \left( \tilde{H}^{on}_{n_i l_i m_i, n'_i l'_i m'_i} + \tilde{H}^{on*}_{n'_i l'_i m'_i, n_i l_i m_i} \right)/2 & i = j \\ H^{off}_{n_i l_i m_i, n_j l_j m_j} = \left( \tilde{H}^{off}_{n_i l_i m_i, n_j l_j m_j} + \tilde{H}^{off*}_{n_j l_j m_j, n_i l_i m_i} \right)/2 & i \neq j \end{cases} \quad (23)$$

## *Tests and applications*

To assess the precision and transferability of HamGNN, we trained and tested HamGNN on the ab initio Hamiltonian matrices and electronic structures for the periodic and aperiodic systems including various molecules, periodical solids, a nanoscale dislocation defect, and a Moiré superlattice. Previously reported models such as SchNorb[33], PhiSNet[34], and DeepH[46] are trained and tested on only one configuration each time, and predicting the Hamiltonian matrix of a different configuration requires additional training on the perturbed structures of that configuration. Since HamGNN is based on the universal parameterized Hamiltonian proposed in this work, our model can be trained and tested on structures with the same atomic species but different configurations in the same way as ML force field models.

### *Molecules*

The QM9 dataset[47,48] contains 134k stable small organic molecules made up of CHONF. These small organic molecules are important candidates for drug discovery. The development of general ML models for rapid screening of the electronic structure properties of drug molecules is beneficial for understanding the mechanisms of drugs and shortening the cycle of drug development. We calculated the real-space ab initio TB Hamiltonian matrices using OpenMX for 10,000 randomly selected molecules from the QM9 dataset. We divided the whole dataset into the training, validation, and test set with a ratio of 0.8: 0.1: 0.1. As shown in Fig. 2(a), the prediction values coincide quite well with the DFT calculated values of the Hamiltonian matrices for the different configurations outside the training set. The mean absolute error (MAE) of the Hamiltonian matrix element predicted by the trained HamGNN model on the test set is only 1.49 meV. Each of the energy levels calculated by the predicted Hamiltonian coincides almost exactly with the DFT-calculated energy levels (see Fig. 2(b)), showing high precision and transferability.

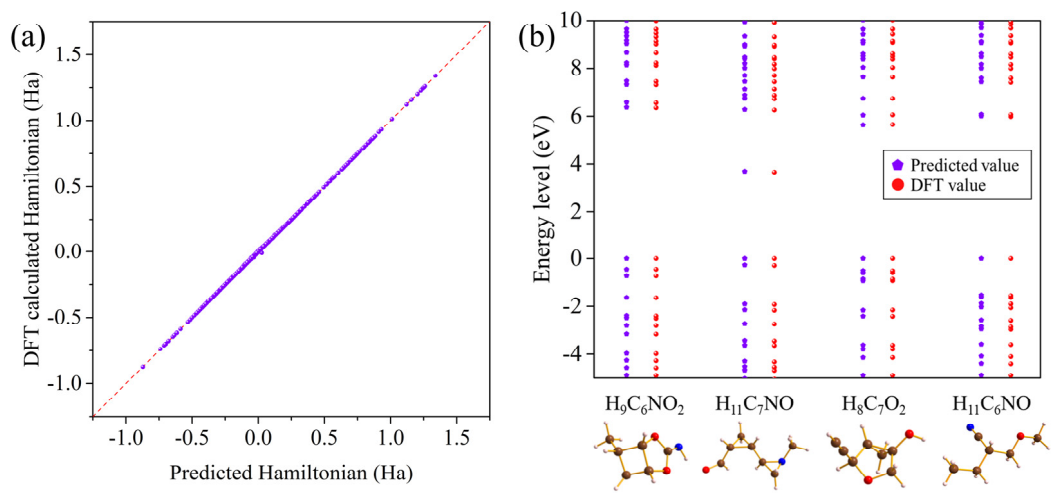

**Fig. 2**. **Application of HamGNN on molecules in the QM9 dataset.** (a) Comparison of the HamGNN predicted Hamiltonian matrix elements with the OpenMX calculated Hamiltonian matrix elements on the QM9 test set. (b) Comparison of predicted and calculated energy levels for 4 molecules randomly selected from the QM9 test set.

We also trained HamGNN on the Hamiltonian of several specific small molecules generated by ab initio molecular dynamics (MD) and compared the accuracy of HamGNN with two recently reported models, PhiSNet[34] and DeepH[46], in predicting the Hamiltonian of these small molecules. The Hamiltonian matrices of these molecules were calculated by OpenMX and divided into the training, validation, and test sets in the same way as PhiSNet in ref.[34]. As can be seen from Table S1, HamGNN achieves the highest accuracy among the models. The accuracy of DeepH is lower than that of PhiSNet and HamGNN because the local coordinate system used by DeepH is not strictly equivariant. Although SE(3) equivariant PhiSNet shows high accuracy in predicting the molecules, it is not a universal equivariant model because it does not satisfy the parity symmetry of the Hamiltonian matrix strictly. Failure will occur in fitting the periodic solid materials containing much more hopping terms or edges (see Appendix A).

*Periodic solids*

We collected 426 carbon allotropes from Samara Carbon Allotrope Database (SACADA)[49], 30 silicon allotropes, and 187 $SiO_2$ isomers from the Materials Project[50]. Each of these structures contains no more than 60 atoms in its unit cell. We performed DFT calculations using OpenMX to obtain the ab initio Hamiltonian matrices for these structures and divided the Hamiltonian matrices in each dataset into training, validation, and test sets with a ratio of 0.8: 0.1: 0.1. The MAEs of the Hamiltonian predicted by HamGNN for the structures in the test set of carbon allotropes, silicon allotropes, and $SiO_2$ isomers are 1.84 meV, 2.60 meV, and 3.75 meV, respectively. The MAE of HamGNN on the carbon allotropes is even lower than the error (2.0 meV) of DeepH on the training dataset of only the graphene structures[46]. Most importantly, our HamGNN model trained on the SACADA dataset is transferable and can fit the Hamiltonian of carbon allotropes of arbitrary sizes and configurations outside the training set.

We used pentadiamond[51], Si (MP-1204046), and $SiO_2$ (MP-667371) to test the accuracy and transferability of the HamGNN models trained on the three datasets. The test structures are shown in Fig. 3(a-c). Pentadiamond is a three-dimensional carbon foam constructed from carbon pentagons and contains 88 carbon atoms in the unit cell[51]. The Si structure labeled MP-1204046 contains 106 atoms in the unit cell and belongs to the tetragonal system. The $SiO_2$ structure labeled MP-667371 is characterized by the complex porous structures built by $SiO_4$ tetrahedra and has 168 atoms in the unit cell. The MAE of the Hamiltonian matrix elements predicted by HamGNN on the pentadiamond, Si (MP-1204046), and $SiO_2$ (MP-667371) is only 2.28 meV, 1.75 meV, and 3.32 meV respectively. The high prediction accuracy can be seen in the scatter plots of the predicted Hamiltonian matrices versus the DFT calculated Hamiltonian matrices shown in Fig. 3(d-f). Using the predicted Hamiltonian, we can obtain the energy bands that almost exactly coincide with those of the DFT calculations, as shown in Fig. 3(g-i). The Hamiltonian matrices and energy bands of twisted bilayer graphene (TBG), Si (MP-1199894), and SiO$_2$ (MP-1200292) are also predicted with similar tiny errors. Details can be seen in Appendix B.

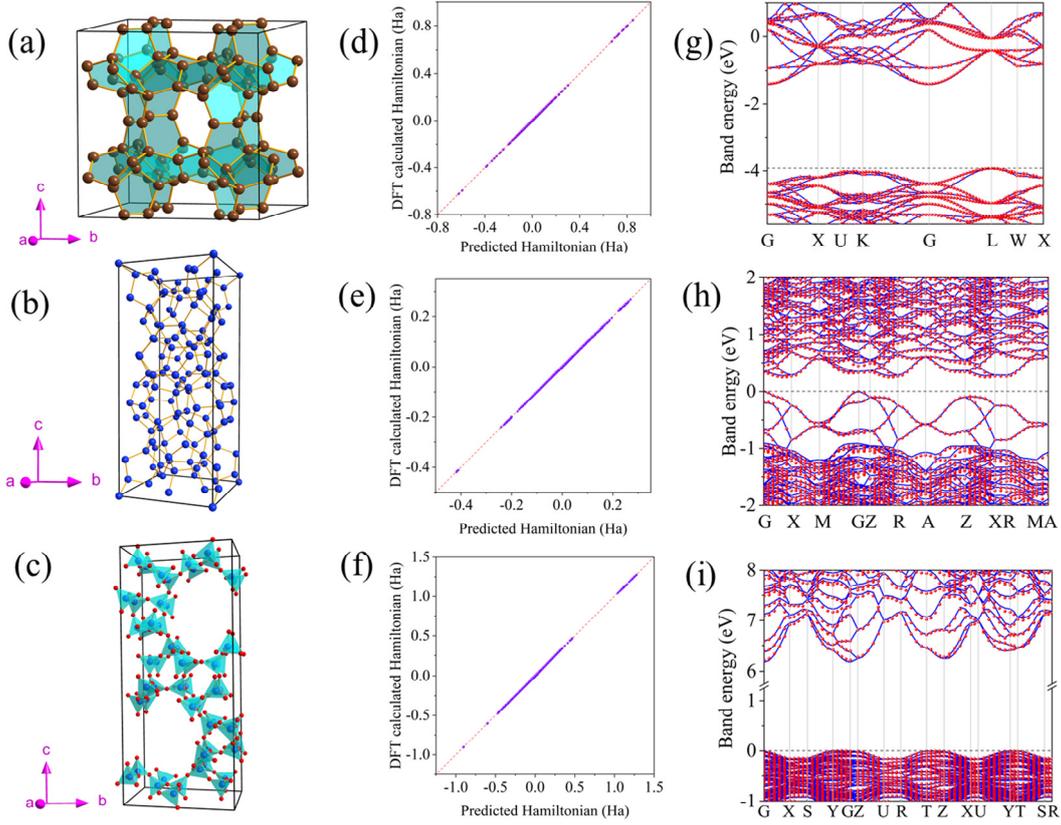

**Fig. 3**. **The prediction of HamGNN on several periodic solids that are not present in the training sets.** (a−c) Crystal structures of pentadiamond, Si (MP-1204046), and $SiO_2$ (MP-667371). (d−f) Comparison of the HamGNN predicted Hamiltonian matrix elements and the DFT calculated Hamiltonian matrix elements of pentadiamond, Si (MP-1204046), and $SiO_2$ (MP-667371). (g−i) Comparison of HamGNN predicted energy bands (solid line) and DFT calculated energy bands (dashed line) of pentadiamond, Si (MP-1204046), and $SiO_2$ (MP-667371).

*Dislocation defect*

The electronic structure of various point defects in semiconductors can be calculated easily by DFT methods[52, 53]. However, the simulation of isolated dislocations requires the use of large supercells to avoid the interactions from the neighbor dislocations, resulting in the high computational cost for traditional DFT to simulate the dislocations accurately. Using the ML model trained in this work to directly fit the ab initio Hamiltonian of a large system can reconcile the computational efficiency with accuracy. In this work, we take an isolated perfect edge dislocation in silicon as an example to demonstrate the application of the HamGNN model to simulate nanoscale defects in large systems. Crystalline silicon has a diamond-type crystal structure, whose most favorable slip system belongs to the type ½⟨110⟩{111}. Taking the {111} plane of silicon as the slip plane, we constructed an isolated edge dislocation with Burgers vector 1/2<110>, as shown in Fig. 4(a). There are 2,428 atoms in this supercell with an isolated edge dislocation. We used the HamGNN model trained on the allotropes of Si to predict the Hamiltonian of this dislocation model and calculated its band structure and the Bloch wavefunction of valence band maximum (VBM) with the predicted Hamiltonian. It can be seen from Fig. 4(b) that an extra occupied energy level appears in the band gap, which greatly narrows the band gap of bulk Si and makes the dislocation to be a one-dimensional metallic chain. The VBM wave function shown in Fig. 4(a) is mainly distributed in the dislocation core, indicating that this occupied level is caused by the hanging bonds and structural distortion in the core. The predicted electronic structure in the dislocation core is similar to that of the DFT simulation for a non-periodical dislocation model containing only 358 atoms[54]. However, the dangling bonds at the edges of the non-periodical supercell are saturated by hydrogens[54], which may introduce unrealistic states. Since HamGNN can build a shortcut from structure to ab initio Hamiltonian matrix, the electronic structure of large supercells can be calculated directly without redundant SCF iterations. HamGNN spent only 467 seconds on a single CPU calculating the Hamiltonian of the supercell containing 2,428 atoms, showing fantastic speed and efficiency.

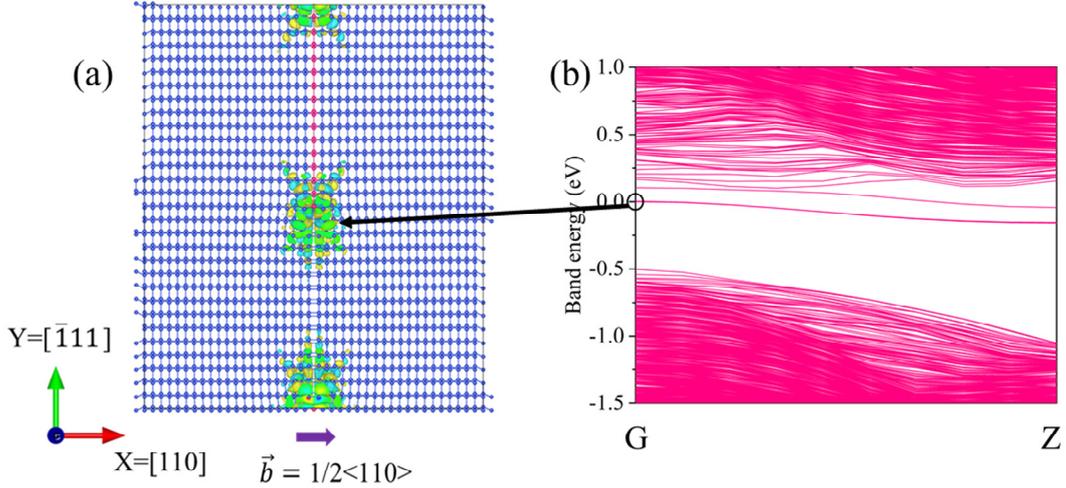

**Fig. 4. The electronic structure prediction on silicon dislocation.** (a) The atomic structure and defect state of an isolated dislocation in silicon. The atoms on the dislocation plane is colored red. (b) The energy bands of the dislocation model along G to Z(0.0, 0.0, 0.5).

*Moiré superlattice of bilayer MoS$_2$*

$MoS_2$ is a 2D transition metal dichalcogenide (TMD) that has attracted much attention because it is an excellent semiconductor with a wide range of applications in the field of electronics and optoelectronics[55-58]. Different from monolayer or untwisted bilayer $MoS_2$, the twisted bilayer $MoS_2$ with Moiré angles has been found to have flat bands and shear solitons[55, 59-61], which could lead to some novel physical phenomena, such as superconducting states, quantum Hall insulators, Mott-insulating phases. HamGNN was trained on the dataset of the untwisted bilayer

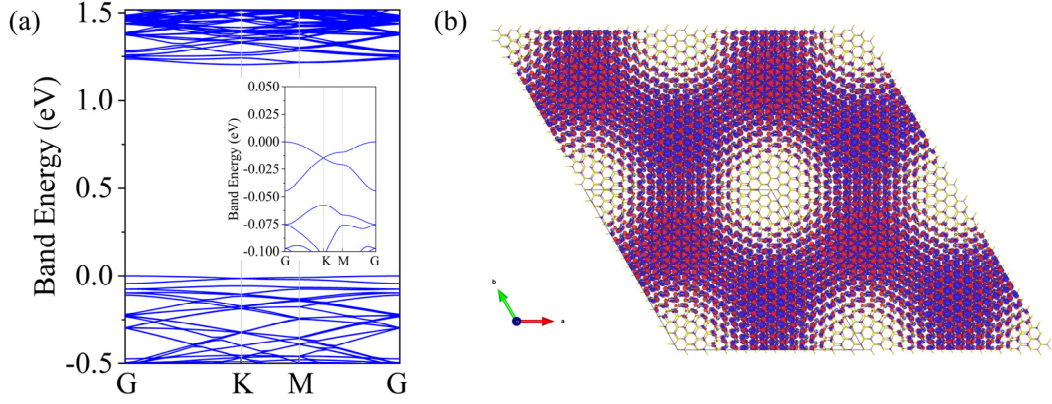

**Fig. 5. The electronic structure prediction on the twisted bilayer MoS$_2$ with a Moiré angle of 3.5°.** (a) The band structure of the twisted bilayer MoS$_2$. (b) The spatial distribution of VBM wave function.

MoS$_2$ structures containing only 54 atoms and then used to predict the electronic structure of twisted bilayer MoS$_2$ superlattice with a Moiré angle. We used OpenMX to calculate the Hamiltonian matrices of 500 MoS$_2$ bilayer structures with a layer slip distance up to 2Å along random directions, and divided the dataset into training, validation, and test sets with a ratio of 0.8: 0.1: 0.1. The MAE of the trained HamGNN on the test set is only 0.52 meV. Fig. 5 shows the calculation results of HamGNN for the electronic structure of the twisted bilayer MoS$_2$ containing 1,626 atoms with a Moiré angle of 3.5°. As shown in Fig. 5(a), the emergence of flat bands at the valence band edge and the Dirac cone at the K point agrees well with the DFT calculations in ref.[59]. Fig. 5(b) shows that the spatial distribution of the predicted VBM wave function is highly localized around the Moiré patterns, which is consistent with the DFT calculations[59]. The traditional DFT methods need to perform a lot of SCF iterations and diagonalization processes to obtain the self-consistent charge density and Hamiltonian matrix, so it takes a lot of time to calculate the large twisted bilayer superlattices. HamGNN can map the structure directly to the ab initio Hamiltonian matrix, and obtain all the information of the electronic structure through only one diagonalization process, which can greatly improve the computational efficiency. The HamGNN model provides an easy way to systematically study the electronic structure of twisted bilayer superlattices with various angles after training on small untwisted bilayer structures.

### *$Bi_xSe_y$ quantum materials*

Bi and Se have multiple chemical valences and can form a set of binary compounds $Bi_xSe_y$ with various stoichiometric ratios[62, 63]. Bi is a heavy element whose d electrons have strong SOC effects. A total of 19 $Bi_xSe_y$ compounds can be found on Materials Project[50]. The compound $Bi_8Se_7$ (id: MP-680214) shown in Fig 6A, which contains 45 atoms in the unit cell, was used to test the transferability and accuracy of HamGNN. The remaining 18 $Bi_xSe_y$ compounds, which contain no more than 40 atoms in the unit cell, were used to generate the training set for the network. To increase the size of the training set, we applied a random perturbation up to 0.02 Å to the atoms of each $Bi_xSe_y$ structure to generate 50 new perturbed structures and obtained 900 structures in total. These structures were randomly divided into the training, validation, and test sets with a ratio of 0.8: 0.1: 0.1. The MAE of the real part of the SOC Hamiltonian predicted by the trained model for $Bi_8Se_7$ is 1.29 meV, and the MAE of the imaginary part of the SOC Hamiltonian is only $5.0\times10^{-7}$ meV. As shown in Fig. 6(b), the predicted and calculated energy bands of $Bi_8Se_7$ is very close. Since the SOC effect is mainly reflected in the imaginary part of the Hamiltonian, such a low MAE for the imaginary part indicates that our proposed parameterized SOC Hamiltonian can describe the SOC effect of different systems very accurately. The training set contains only 18 perturbed structures of $Bi_xSe_y$ compounds and no compounds with the stoichiometric ratio of $Bi_8Se_7$ are present, but the HamGNN model still accurately predicted the SOC Hamiltonian and the energy band of this structure, showing very high transferability.

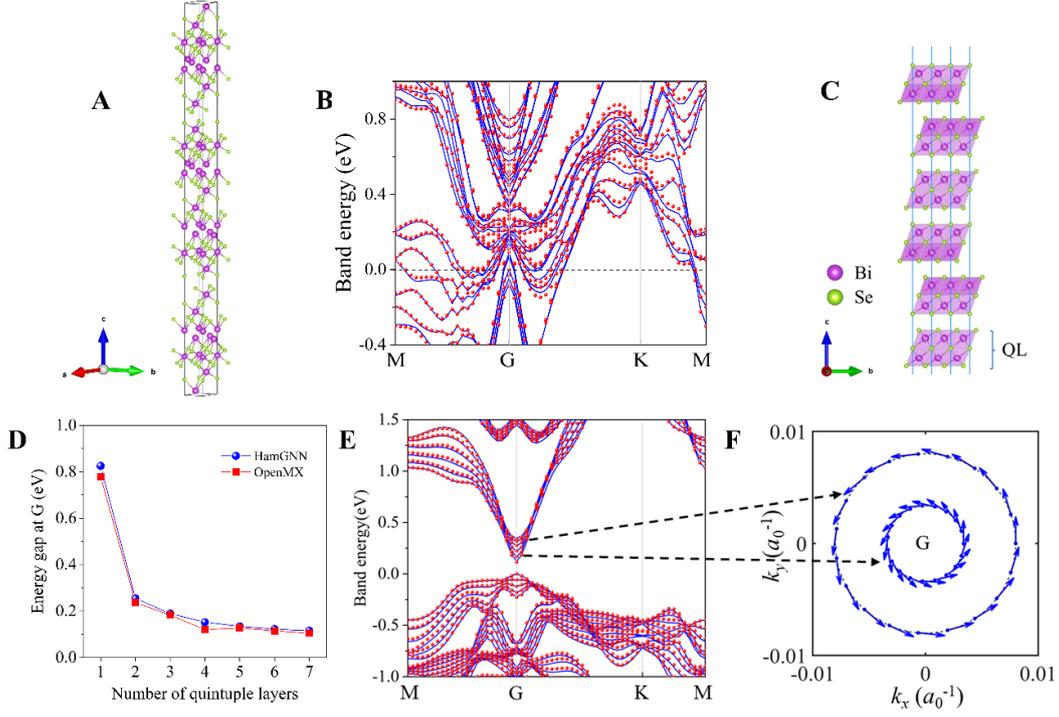

**Fig. 6. The electronic structure prediction on the $Bi_xSe_y$ quantum materials.** (a) The crystal structure of $Bi_8Se_7$. (b) Comparison of HamGNN predicted energy bands (solid line) and DFT calculated energy bands (dashed line) of $Bi_8Se_7$. (c) Schematic diagram of the layered crystal structure of $Bi_2Se_3$. (d) Comparison of HamGNN prediction and DFT calculations of the energy gap at G point. (e) Comparison of HamGNN predicted energy bands (solid line) and DFT calculated energy bands (dashed line) of $Bi_2Se_3$ with 6 QLs. (f) The predicted spin textures on the lowest unoccupied state of 0.07 eV and 0.23 eV above the conduction band minimum (CBM).

$Bi_2Se_3$ is a well-known 3D topological insulator material and is a good platform to study the quantum effects related to SOC effects[64-68]. The bulk $Bi_2Se_3$ is an insulator, while a metallic state protected by the time-reversal symmetry is formed on the surface. Bulk $Bi_2Se_3$ is stacked by quintuple layers (QLs) through the van der Waals (vdW) interaction, as is shown in Fig. 6(c). Each QL layer is composed of five atomic layers of Se-Bi-Se-Bi-Se combined by strong covalent bonds. The HamGNN predicted and DFT calculated G-point band gaps of the $Bi_2Se_3$

Slab model with 1 to 7 QLs are very close, as shown in Fig. 6(d). It can be seen from Fig. 6(e) that the slab model with a single QL has the largest G-point band gap. When a new QL layer is added to the slab, the G-point band gap decreases rapidly under the influence of van der Waals interactions. As the number of QL layers increases, $E_g(G)$ gradually decreases, and the band dispersion at the G point gradually tends to be linear to form a Dirac cone. As shown in Fig. 6(e), A Dirac cone with a small gap appears near the Fermi surface at the G point. The spin textures on the lowest unoccupied state with 0.07 eV and 0.23 eV above conduction band minimum (CBM) were calculated using the HamGNN predicted Hamiltonian matrix, as is shown in Fig. 6(f). The predicted spin textures are in good agreement with the fact that the Dirac cone is a topological surface state protected by time-reversal symmetry and that spin and momentum on the topological surface state are bound.

## Discussion

DFT methods are now widely used to calculate various properties of molecules and materials. However, successful DFT calculations on large systems are still rare because of the prohibitive computational resources and running time required. A typical DFT calculation often requires tens to hundreds of self-consistent iterations to obtain the final Hamiltonian and wave function, and the diagonalization of the Hamiltonian on a dense k-point grid is carried out in each iteration step. This process takes up most of the running time of DFT calculation and can not be skipped. In recent years, the emergence of deep learning enables efficient atomic simulations with DFT accuracy. Machine learning force fields (MLFFs) with quantum mechanical precision are now widely used to accelerate long-time molecular dynamics simulations of large systems. Since potential energy is just an invariant scalar, the implementation of universal MLFF models is relatively easy. While the Hamiltonian is a matrix with rotational equivariance and parity symmetry, the implementation of a transferrable model for directly predicting the Hamiltonian is very difficult.

In this work, an analytical E(3) equivariant parameterized Hamiltonian that explicitly takes into account rotation equivariance and parity symmetry is proposed and further extended to a parameterized Hamiltonian satisfying SU(2) and time-reversal equivariance to fit the Hamiltonian with SOC effects or ionic magnetic moments. Based on this parameterized Hamiltonian, we develop an E(3) equivariant deep neural network called HamGNN to fit the Hamiltonian of arbitrary molecules and solids. Previously reported models were trained and tested on the datasets of the molecular dynamics perturbed molecules and solids with just the same configuration. To demonstrate the accuracy and transferability of this parameterized Hamiltonian, we used the trained HamGNN model to predict the electronic structures of the molecules, periodic solids, the silicon dislocation defect, Moire bilayer $MoS_2$, and $Bi_xSe_y$ quantum materials. Actual tests show that our model has a high accuracy compared with DFT and a high transferability similar to the machine learning force field. These features are the important foundation for the wide application of machine learning electronic structure methods. Since our model can establish a direct mapping from the structure to the self-consistent Hamiltonian without the time-consuming self-consistent iterative process in DFT, it can be used to accelerate the electronic structure calculation of large systems and other costly advanced calculations, such as the electron-phonon coupling matrix via the automatic differentiation ability of the neural network.

## Methods

*Hamiltonian datasets*

QM9 structure set and the molecules in Table S1 are available from http://quantum-machine.org/datasets/, SACADA structure set is available from https://www.sacada.info/sacada_3D.php, and the structures of Si allotropes, $SiO_2$ isomers, and $Bi_xSe_y$ crystals are downloaded from the Materials Project site[69]. To prepare the training set of untwisted bilayer $MoS_2$, a random perturbation of up to 0.02 Å is applied to each atom and a slip of up to 2Å is performed in a random direction within the XY plane. We performed DFT calculations on the structures in the above datasets to obtain TB Hamiltonian matrices via OpenMX[17], a software package for nano-scale material simulations based on norm-conserving pseudopotentials and pseudo-atomic localized basis functions. H6.0-*s*2*p*1, C6.0-*s*2*p*2*d*1, N6.0-*s*2*p*2*d*1, O6.0-*s*2*p*2*d*1, F6.0-*s*2*p*2*d*1, Si7.0-*s*2*p*2*d*1, Mo7.0-*s*3*p*2*d*2, Bi8.0-*s*3*p*2*d*2, and Se7.0-*s*3*p*2*d*2 pseudoatomic orbitals (PAOs) were used as the basis for the calculations. The truncation radius of the atomic orbits of H, C, N, O, and F is 6.0 Bohr, the truncation radius of the atomic orbits of Si, Mo, and Se is 7.0 Bohr, the truncation radius of the atomic orbits of Bi is 8.0 Bohr. The Si dislocation model was built by Atomsk[70] and relaxed by GPUMD[71] with a force criterion of 0.01 eV/Å. Moiré superlattice of bilayer $MoS_2$ was relaxed using VASP[72, 73] with DFT-D2 correction for vdW interaction.

*Network and training details*

The Pytorch-1.11.0 [74], PyG-2.0.4 [75], Pymatgen [76], Nequip-0.5.3 [31], and e3nn-0.5.0 [77] libraries are used to implement HamGNN. All models used in this work have five orbital convolution layers, a pair interaction layer, an on-site layer, and an off-site layer. The edge distances between each atom and its neighbors are expanded by eight Bessel function bases. The spherical harmonic functions with a maximum degree $L_{max} = 4$ are used to embed the directions of the edges. The O(3) representations used for the atom features and pair interaction features have $N_{fea}$ channels with a maximum degree $L_{max}$ and parities p = ±1. The number of feature channels $N_{fea}$ and the maximum degree $L_{max}$ for each dataset are listed in Table S2. A two-layer MLP with 64 neurons is used to map the invariant edge embeddings to the weights of each tensor product path in Eqs. 17 and 19. Shifted softplus [29] function is used as the activation function in the MLP. The gate activation function scales the input features $\left(\bigoplus_i u_{p_i}^{(0)}\right) \oplus \left(\bigoplus_j v_{p'_j}^{(0)}\right) \oplus \left(\bigoplus_j w_{p_j}^{(l_j > 0)}\right)$ with its invariant field $\left(\bigoplus_i u_{p_i}^{(0)}\right)$ and $\left(\bigoplus_j v_{p'_j}^{(0)}\right)$ as the gate. The output equivariant features of gate nonlinearity are $\left[\bigoplus_i \phi_{p_i}^1\left(u_{p_i}^{(0)}\right)\right] \oplus \left[\bigoplus_j \phi_{p'_j}^2\left(v_{p'_j}^{(0)}\right) w_{p_j}^{(l_j)}\right]$, where $\phi_{p_i}^1$ and $\phi_{p'_j}^2$ are the activation functions that vary with the parity of the scalar input, defined as follows[31]:

$$\phi_{p_i}^1(x) = \begin{cases} ssp(x) = \ln(0.5 \times e^x + 0.5) & p_i = 1 \\ \tanh(x) = \dfrac{e^x - e^{-x}}{e^x + e^{-x}} & p_i = -1 \end{cases} \quad (24)$$

$$\phi_{p'_j}^2(x) = \begin{cases} ssp(x) = \ln(0.5 \times e^x + 0.5) & p'_j = 1 \\ |x| & p'_j = -1 \end{cases} \quad (25)$$

$\phi_{p_i}^1(x)$ has the same parity as the input scalar $x$, while $\phi_{p'_j}^2(x)$ always has even parity. This ensures that the parity of the output features of the Gate activation function is equivariant.

To increase transferability and avoid overfitting, we include the error of the calculated energy bands as a regularization term in the loss function:

$$L = \left\|\tilde{\mathbf{H}} - \mathbf{H}\right\| + \frac{\lambda}{N_{orb} \times N_k} \sum_{k=1}^{N_k} \sum_{n=1}^{N_{orb}} \left\|\tilde{\varepsilon}_{nk} - \varepsilon_{nk}\right\| \quad (26)$$

where the variables marked with a tilde refer to the corresponding predictions and $\lambda$ denotes the loss weight of the band energy error. $\lambda$ equals 0.001 in our training. When the training of the network has not converged, the error of the predicted Hamiltonian is large, resulting in poor or even divergent prediction values of the energy bands. Adding the band loss value at the beginning of training may cause the total loss value to diverge. Therefore, we train the network in two steps. First, only the mean absolute error of Hamiltonian matrices is used as the loss value to train the network until the network weights converge. The parameters were optimized with AdamW[78, 79] optimizer using an initial learning rate of $10^{-3}$. Then the mean absolute error of each band calculated at $N_k$ random points in the reciprocal space is added to the loss function and starts the training at an initial learning rate of $10^{-4}$. When the accuracy of the model on the validation set is not improved after successive $N_{patience}$ epochs, the learning rate will be reduced by a factor of 0.5. When the accuracy of the model on the validation set is not improved after successive $N_{stop}$ epochs or the learning rate is lower than $10^{-6}$, the training will be stopped and the model that has the best accuracy on the validation set will be used on the test set. The values of some key network and training parameters on each dataset are listed in Table S2. All models were trained on a single NVIDIA A100 GPU.

## Acknowledgments

We thank Dr. Hongli Guo for providing the structure of the Moiré superlattice of bilayer $MoS_2$. We acknowledge financial support from the Ministry of Science and Technology of the People´s Republic of China (No. 2022YFA1402901), NSFC (grants No. 11825403, 11991061, 12188101), and the Guangdong Major Project of the Basic and Applied Basic Research (Future functional materials under extreme conditions--2021B0301030005).

## Data and Code availability

The additional data and code that support the results of this work can be requested from the corresponding author after the manuscript is accepted.